\documentclass[amsmath,aps,10pt,twocolumn,superscriptaddress]{revtex4} 

\usepackage{amssymb,url}
\usepackage{graphicx}
\usepackage{hyperref}

\usepackage{array}
\usepackage{amsmath}

\long\def\del #1 \enddel { }

\usepackage{graphicx}
\usepackage{amsmath}
\usepackage{amssymb}
\usepackage{subfigure}

\usepackage{epsfig}

\usepackage{graphicx}
\usepackage{subfigure}
\usepackage{hyperref}

\def\beq{\begin{equation}}
\def\eeq{\end{equation}}

\def\bea{\arraycolsep .1em \begin{eqnarray}}
\def\eea{\end{eqnarray}}

\def\eq#1{(\ref{#1})}

\def\s0#1#2{\mbox{\small{$ \frac{#1}{#2} $}}}
\def\0#1#2{\frac{#1}{#2}}

\def\grgl{\:\hbox to -0.2pt{\lower2.5pt\hbox{$\sim$}\hss}{\raise3pt\hbox{$>$}}\:}
\def\klgl{\:\hbox to -0.2pt{\lower2.5pt\hbox{$\sim$}\hss}{\raise3pt\hbox{$<$}}\:}

\begin{document}

%\title{Bootstrap approach for asymptotically safe gravity}
\title{A bootstrap strategy for asymptotic safety}
\author{K.~Falls}
\author{D.F.~Litim}
\author{K.~Nikolakopoulos}
\affiliation{Department of Physics and Astronomy, University of Sussex, Brighton, BN1 9QH, U.K.}
\author{C.~Rahmede}
\affiliation{\mbox{Institut f\"ur Physik, Technische Universit\"at Dortmund, 
44221 Dortmund, Germany}}
\affiliation{\mbox{Institute for Theoretical Physics,  Karlsruhe Institute of Technology,
76128 Karlsruhe, Germany}}

\begin{abstract}A search strategy for
asymptotic safety  is put forward and  tested  for a simplified version of 
gravity in four dimensions using the renormalization group. 
Taking the  action to be a high-order polynomial of the Ricci scalar, 
a self-consistent ultraviolet fixed point is found 
where curvature invariants become increasingly irrelevant with increasing mass dimension. Intriguingly,  universal scaling exponents take near-Gaussian values despite the presence of residual interactions. 
Asymptotic safety of metric gravity would seem in reach if this pattern carries over to the full theory.

\end{abstract}

\maketitle

\section{Introduction}

An intriguing consequence of quantum physics is that the  strength of 
fundamental forces 
depends on the energy 
scale at which they are probed. The energy-dependence of couplings then teaches us how quantum fluctuations modify the laws of physics when moving from largest to smallest energies.  It is commonly believed that  couplings should approach a fixed point at highest energies for a quantum field theory to exist fundamentally \cite{Wilson:1970ag}. Such theories are then said to be either asymptotically free, if  the fixed point is of a non-interacting type, or asymptotically safe \cite{Weinberg:1980gg}, otherwise.

A role model for a fundamental and 
asymptotically free theory in four dimensions is quantum chromodynamics (QCD), the theory of the strong nuclear force. It describes the interaction of colored matter fields, the quarks, with the carrier of the nuclear force, the gluons. Classically, the coupling constant of the strong force $\alpha_s$ is marginal. Quantum-mechanically, the fluctuations of the quark and the gluon fields turn the coupling into an energy-dependent running coupling. In the high-energy limit, QCD is weakly coupled and
approaches a free ultraviolet fixed point $\alpha_s\to 0$.
Asymptotic freedom 
has the additional benefit that canonical power counting is applicable. The knowledge that only a few relevant or marginal invariants persist at high energies is at the root of reliable approximations and predictions within 
continuum and lattice studies of QCD.

A candidate for a fundamental and asymptotically safe theory in four dimensions is gravity \cite{Weinberg:1980gg}. 
It describes the interactions of matter with the carrier of the gravitational force, the metric field. Classically, Newton's coupling is a constant. Quantum-mechanically, metric fluctuations
give rise to an energy-dependence of couplings and, possibly, a fixed point at high energies. Similarly to QCD, this would then provide a link between the high- and the low-energy regimes by following the renormalization group (RG) evolution of couplings. 
On the other hand, and very much unlike QCD, the metric couplings are expected to reach an interacting rather than a free high energy  fixed point.
This innocuous-looking difference between asymptotic freedom and asymptotic safety has the profound implication that the set of relevant and marginal invariants is no longer known beforehand. The asymptotic safety conjecture thus seems to require novel search strategies beyond perturbation theory.

In this Letter we put forward a self-consistent strategy to test the asymptotic safety conjecture. Our approach compensates the lack of beforehand information with the help of a bootstrap. Thereby, it also gives access to the 
scaling dimensions of the theory at a fixed point, which are worked out for the example of $f(R)$-type theories of gravity. 
We focus on the main ideas and results, and leave technicalities for a seperate publication
\cite{Falls:2014tra}.

\section{Asymptotic safety}

We recall the asymptotic safety conjecture following the ideas of S.~Weinberg for the example of gravity \cite{Weinberg:1980gg}. To that end, we consider a Wilsonian  effective action for four-dimensional Euclidean gravity of the form
\begin{equation}\label{Gammak}
\Gamma_k=
\,\sum_i\bar\lambda_i
\,\int d^4x
\,{\cal O}_i 
\end{equation}
where the terms ${\cal O}_i$ are
built out of the metric field and its derivatives in accordance with diffeomorphism invariance,
and $\bar\lambda_i$ are the corresponding couplings. 
The couplings and the effective action also depend on the RG momentum scale $k$. Under the RG evolution of parameters,   the effective action \eq{Gammak} should approach the Einstein-Hilbert action at low energies $(k\to 0)$ to recover classical general relativity, 
\beq\label{S}
\Gamma\approx \int d^4x \sqrt{g} \left[\frac{\Lambda}{8\pi\,G}-\frac{R}{16\pi G}\right]\,,
\eeq
where $G\approx6.67 \times  10^{−11}\,{\rm  m}^3/({\rm kg\, s}^2)$ is Newton's constant, ${\Lambda}/(8\pi\,G)\approx 10^{-26} {\rm kg}/{\rm m}^3$ the vacuum energy, $R$ the Ricci scalar, and   $g=\det g_{\mu\nu}$. 
At high energies $(k\to \infty)$, it is assumed that
$\Gamma_k$ approaches an asymptotically safe fixed point $\Gamma_*$. In its vicinity, gravity remains interacting yet the running Newton coupling  $G_k$
becomes very small compared to its classical value $G_k\ll G$ \cite{Litim:2011cp}. 
An asymptotically safe RG trajectory $k\to\Gamma_k$ then connects the ultraviolet and infrared limits of gravity inasmuch as it does for asymptotically free theories, 
eg.~QCD. 

In addition, the feasibility of asymptotic safety also hinges on the residual interactions at the putative fixed point, which can be understood as follows \cite{Weinberg:1980gg}. Denoting the canonical mass dimension of the couplings as $[\bar\lambda_i]=d_i$, we introduce dimensionless couplings
$\lambda_i=k^{-d_i}\bar\lambda_i(k)$. If the  term ${\cal O}_i$ in the effective action \eq{Gammak} contains $2m_i$ derivatives of the metric field, we have
$d_i=4-2m_i$.
Note that we adopt conventions where the metric field is dimensionless, and  derivatives 
have mass dimension one, but
our conclusions will be independent thereof. In these conventions,  
the RG scale dependence  of couplings is given by $\beta_i\equiv k\partial_k \lambda_i$.
The dimensionless couplings become independent of the RG scale $\lambda_i=\lambda_i^*$ if $\beta_i=0$ and the theory is said to be at a fixed point.
Linearizing in its vicinity,
we find
\begin{equation}\label{beta}
\beta_i=\sum_j M
_{ij}\,(\lambda_j-\lambda_j^*)+{\rm subleading}
\end{equation}
where $M
_{ij}=\left.\partial\beta_i/\partial\lambda_j\right|_*$ is  the stability matrix.
Integrating \eq{beta}, the general solution takes the form
\begin{equation}\label{lambda}
\lambda_i(k)=\lambda_i^*+\sum_n \,c_n\, V^n_i\, k^{\vartheta_n}+{\rm subleading}
\end{equation}
 where $\vartheta_n$ are the  eigenvalues of $
M$, $V^n$ the corresponding eigenvectors, and $c_n$ free parameters. The significance of \eq{lambda} is as follows: 
In order to have a well-defined  ultraviolet limit the irrelevant eigendirections 
which would diverge for $k\to \infty$, i.e.~those with  positive $\vartheta_n>0$, must vanish, implying that
the coefficients
$c_n=0$. This condition defines the ultraviolet critical surface of asymptotically safe gravity.
The relevant eigendirections with $\vartheta_n<0$ are unconstrained and the corresponding coefficients $c_n$ are free parameters of the theory. Provided that the number of relevant directions  with $\vartheta_n<0$ remains finite, this leads to a predictive theory with 
a finite number of free parameters whose values should be fixed by experiment.  
We note that since the stability matrix $M$ is real but not manifestly symmetric, some eigenvalues may come out as complex conjugate pairs. Then the sign of the real part determines whether the associated scaling fields are relevant. We conclude from this discussion that the set of eigenvalues $\{\vartheta_n\}$, and in particular the subset of relevant and marginal ones, are a crucial ingredient for asymptotically safe theories.

\section{The bootstrap}

For asymptotically free theories, 
canonical power counting fixes the set of eigenvalues $\{\vartheta_n\}$ without further ado,
and their values can be read off from \eq{beta} in the weak coupling limit \cite{Weinberg:1980gg}
\beq\label{DA}
\beta_i=-d_i\,
\lambda_i+{\rm fluctuations}\,.
\eeq
The fluctuation-induced terms involve higher powers in  the couplings and  the set of Gaussian eigenvalues 
are  (minus) the canonical mass dimension of the  couplings 
$\{\vartheta_{{\rm G},n}=-d_n\}$. 
For asymptotically safe theories, residual interactions modify the eigenvalues.  To 
test the asymptotic safety conjecture, a  strategy is required which reliably determines  
 the fixed point -- provided it exists -- and the set of scaling exponents $\{\vartheta_n\}$. Unlike in asymptotically free theories, it is not known 
{\it a priori}  
which invariants will turn out to be relevant, marginal, or irrelevant at a fixed point. As such, the lack of beforehand information may hamper the reliability of fixed point studies, as these often have to adopt  some approximation for \eq{Gammak}.
On the other hand, it is conceivable that the residual interactions only change the sign of, at best, finitely many eigenvalues \cite{Weinberg:1980gg}. If so,  the relevancy of invariants at an interacting fixed point would continue to be governed by their canonical mass dimension.

In this light, it is tempting to compensate  the absence of prior knowledge  by an iterative strategy:
Under the 
working hypothesis that invariants with a larger canonical mass dimension remain  less relevant even in the interacting case, the action \eq{Gammak} is approximated by a few terms ${\cal O}_i$ up to a pre-set maximum canonical mass dimension $D$, say $[{\cal O}_i]\le D$ (Step 1).
The impact of interactions on the universal scaling exponents is then quantified using suitable methods
eg.~Wilson's RG in the continuum, 
lattice simulations, 
or holography  
(Step 2). Within the  RG, 
this amounts to the explicit derivation of RG flows for all couplings, the search for interacting ultraviolet fixed point(s) in terms of these, and the computation of the corresponding set of  non-perturbative eigenvalues $\{\vartheta_n\}$. 
Finally, we 
increase the pre-set maximum $D$ -- and thus the set of invariants and couplings retained in \eq{Gammak} --  and re-do the previous analysis
(Step 3).  In fact, the iteration of Step 2 and 3 is crucial as it offers a systematic  control 
of which invariants are responsible for the values of scaling exponents, and whether their magnitudes continue to be governed by the canonical mass dimension. If the initial assumption is confirmed to sufficiently high order,  the self-consistency of the working hypothesis 
is validated {\it a posteriori}.
If the strategy fails even to high orders in the iteration, 
and it may do so in a variety of manners, 
the working hypothesis is refuted.

\section{Quantum gravity}

We now apply our strategy to a simplified version of quantum gravity also relevant for cosmology  \cite{DeFelice:2010aj}, where the action \eq{Gammak} is taken to be a local function of the Ricci scalar only. 
In Step 1  we  specify the terms ${\cal O}_n$ in \eq{Gammak}
as powers of the Ricci scalar $\sqrt{g}R^n$,
\begin{equation}\label{GammaR}
\Gamma_k=
\,\sum_{n=0}^{N-1}
\lambda_n
\,k^{d_n}
\,\int d^4x\sqrt{g}
\,R^n\,.
\end{equation}
The Ricci scalar contains two derivatives of the metric field and the Gaussian exponents in this theory are 
\beq\label{G}
\{\vartheta_{{\rm G},n}=2n-4,\ 0\le n\le N-1\}\,.
\eeq
Classically the theory has two relevant and a marginal coupling associated to the vacuum energy, the Ricci scalar,
and the $R^2$ term. The canonical  mass dimensions of the terms in \eq{GammaR} are bounded by $[{\cal O}_n]\le D=2(N-1)$.  
For Step 2, we adopt Wilson's
RG \cite{Wetterich:1992yh}
for the gravitational action \eq{Gammak}. The RG flows \eq{DA} for all couplings in \eq{GammaR} 
are obtained using techniques developed in 
\cite{Reuter:1996cp,Litim:2001up,Litim:2003vp,Codello:2008vh,Falls:2014tra}, also adding suitable gauge-fixing and ghost terms within the background field formalism in
the conventions of 
\cite{Codello:2008vh}.
We then fix $N$, identify the interacting fixed points, and compute the set of $N$ universal eigenvalues
\beq\label{Theta}
\{\vartheta_n(N), \ 0\le n\le N-1\}\,
\eeq
which characterise the critical theory. The eigenvalues are ordered according to magnitude (of their real parts, if complex). In Step 3 we then increase $D$, which is equivalent to increasing  $N\to N+1$, perform a new fixed point search, compute the scaling exponents \eq{Theta}, and compare with results at lower $N$.

We have iterated our procedure from $N=2$ to $N=35$. 
A numerically stable fixed point is found to each and every order. Earlier results up to  order $N=9$ \cite{Codello:2008vh} and $N=11$ \cite{Bonanno:2010bt} are reproduced, and serve as an important  consistency check. Occasionally, a spurious fixed point is found at a few specific orders, which is discarded.
\begin{figure}[t]
\centering
\begin{center}
\unitlength0.001\hsize
\begin{picture}(1000,1600)
\put(0,0){\includegraphics[width=\hsize]{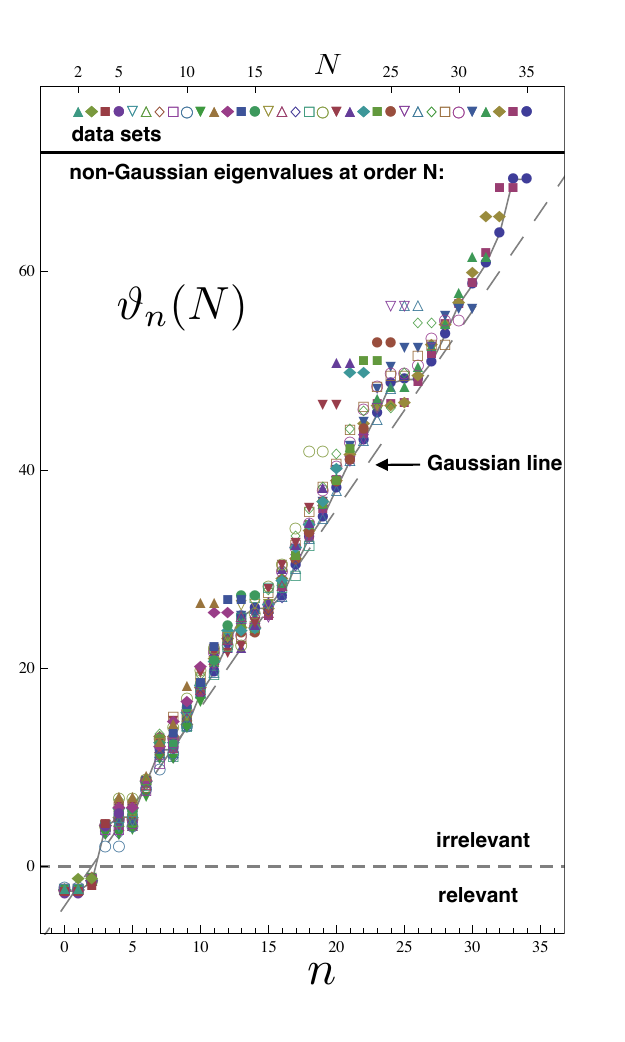}}
\end{picture}
\end{center}
\vskip-.8cm
\caption{\label{ScalingW2} The complete sets of eigenvalues at the ultraviolet fixed point  \eq{Theta}
for all  $N$, sorted by magnitude.
The results at the highest order $(N=35)$ are linked by a line to guide the eye. The long dashed line indicates Gaussian scaling. The inset (upper panel) relates the data sets  at approximation order $N$ with the symbols used in the lower panel.}
\end{figure} 
The stable fixed point converges rapidly, and is used to compute the scaling exponents \eq{Theta} for all $N$.
Our results 
are shown in Fig.~\ref{ScalingW2} for 34 consecutive orders $N$.
We find that the theory has three relevant (negative) eigenvalues with $\vartheta_n<0$. For fixed $n$, we note a good convergence of $\vartheta_n(N)$ with increasing approximation order $N$. 
We also note the occasional appearance of complex conjugate pairs of eigenvalues, in which case we have plotted their real parts. Complex eigenvalues hint towards a degeneracy induced by strong correlations. In a more complete treatment these degeneracies may be lifted through  interactions beyond those retained here, 
eg.~Weyl tensor invariants \cite{Niedermaier:2009zz,Benedetti:2009rx} or dynamical ghosts \cite{Christiansen:2012rx}.
Note also that if the largest eigenvalues $\vartheta_{N-1}(N)$ and  $\vartheta_{N-2}(N)$  per data set \eq{Theta} are a complex conjugate pair, their real parts tend to deviate more strongly from their asymptotic values. The reason for this is that the values of the largest exponents per approximation order are sensitive to the couplings neglected at order $N$, eg.~$\lambda_{N}$ or higher. This pattern is well-known from analogous studies of critical scalar theories \cite{Litim:2002cf}. In fact, as soon as the next few couplings are taken into account, the exponent converges well, see Fig.~\ref{ScalingW2}.  To conclude, the main result of Fig.~\ref{ScalingW2} is that the ordering of scaling exponents \eq{Theta} is indeed controlled by the underlying canonical dimension even at an interacting fixed point, thus supporting the
working hypothesis.

\section{Near-Gaussianity}

We now turn to the large-order behavior of scaling exponents. Since
the eigenvalues appear to be growing linearly with $n$,  
we perform a least-square fit of \eq{Theta} as
\beq\label{shift}
\vartheta_n
\approx a\cdot n-b
\eeq 
for 24 data sets with $12\le N\le 35$. 
For each fit, we omit the two largest values in \eq{Theta} for reasons detailed above. 
The correlation coefficients are very close to one for the fits of all data sets, and
the non-perturbative coefficients \eq{shift} at the ultraviolet fixed point read
\begin{equation}
\label{abUV}
\begin{array}{rcl}
a_{\rm UV}&=&2.17\pm 5\% \\[.5ex]
b_{\rm UV}&=&4.06\pm10\%\,,
\end{array}
\end{equation}
where the error estimate, roughly a standard deviation, arises from the average over data sets  \cite{Litim:2010tt}.    The smallness of the estimated error in 
$a_{\rm UV}$ permits an extrapolation of the result \eq{shift}, \eq{abUV} towards higher $n$.
We therefore have good reason to expect that higher order terms
such as $\sqrt{g}R^{256}$ continue to be irrelevant 
 as a consequence of residual interactions.

The differences between \eq{abUV} and the Gaussian coefficients $a_{\rm G}=2$ and $b_{\rm G}=4$ serve as an indicator for the non-perturbative corrections due to asymptotically safe interactions.
Interestingly, our results 
show that the scaling exponents \eq{Theta} remain near-Gaussian. 
The off-set $b_{\rm UV}$ is compatible with the classical value, though with a slight bias towards larger values, whereas the slope $a_{\rm UV}$ comes out larger than the Gaussian slope. 
It is tempting to speculate that this may be a consequence of the smallness of Newton's coupling at an ultraviolet fixed point, though more work is required to clarify this aspect.

\section{Discussion}
We have put forward a  strategy to test the asymptotic safety conjecture by means of a 
 bootstrap to  compensate for the {\it a priori} unknown scaling exponents.
The price to pay is the necessity for a  non-perturbative analysis of the fixed point at each and every order in the iteration. Unfortunately, it is not  guaranteed that the procedure succeeds. If it does, however,  the benefits would seem plentiful, including the prospect of a systematic control towards asymptotically safe quantum field theories. 

The procedure has been applied for a simplified version
of quantum gravity. Up to very high order, the model displays a stable ultraviolet fixed point and a self-consistent set of universal eigenvalues.  
The main qualitative effect of residual interactions is the occurence of an additional negative eigenvalue in the spectrum, $\vartheta_2$. The remaining scaling exponents settle for near-Gaussian values.
If a structural reason for this were to 
be found, it could offer a perturbative access towards
 the fixed point \cite{Niedermaier:2009zz}. 

One may speculate that asymptotic safety of metric gravity is in reach provided that the  pattern carries over to the full theory. 
However, much more work needs to be done before definite conclusions are achieved.
Additional  insights may arise from other non-perturbative 
techniques besides the RG, such as the lattice
\cite{Laiho:2011ya,Ambjorn:2011cg,Hamber:1999nu}  or the powerful machinery of holography \cite{Litim:2011qf}, to put 
asymptotic safety to the test.\\[-2ex]

{\it Acknowledgements.}\ We thank H.~Nicolai for discussions. This work is supported by the Science Technology and Facilities Council (STFC) under grant number [ST/J000477/1], and by the A.S.~Onassis Public Benefit Foundation under grant number [F-ZG066/2010-2011].

\end{document}